\pdfoutput=1

\documentclass[11pt,twoside,a4paper,cmspaper,final,collab]{cms-tdr}
\def\svnVersion{311163}\def\cmsCernNo{2015-001}\def\cmsCernDate{\today}\def\cmsMessage{Submitted to Physics Letters B}

\begin{document}\cmsNoteHeader{EXO-12-057}

\hyphenation{had-ron-i-za-tion}
\hyphenation{cal-or-i-me-ter}
\hyphenation{de-vices}
\RCS$Revision: 288969 $
\RCS$HeadURL: svn+ssh://alverson@svn.cern.ch/reps/tdr2/papers/EXO-12-057/trunk/EXO-12-057.tex $
\RCS$Id: EXO-12-057.tex 288969 2015-05-15 02:21:38Z alverson $
\newlength\cmsFigWidth
\ifthenelse{\boolean{cms@external}}{\setlength\cmsFigWidth{0.98\columnwidth}}{\setlength\cmsFigWidth{0.8\textwidth}}
\newlength\cmsFigWidthThree
\ifthenelse{\boolean{cms@external}}{\setlength\cmsFigWidthThree{0.98\columnwidth}}{\setlength\cmsFigWidthThree{0.6\textwidth}}
\ifthenelse{\boolean{cms@external}}{\providecommand{\cmsLeft}{top}}{\providecommand{\cmsLeft}{left}}
\ifthenelse{\boolean{cms@external}}{\providecommand{\cmsRight}{bottom}}{\providecommand{\cmsRight}{right}}
\newcommand{\VmN}{\ensuremath{V_{\mu \mathrm{N}}}\xspace}
\newcommand{\mN}{\ensuremath{m_\mathrm{N}}\xspace}
\newcommand{\N}{\ensuremath{\mathrm{N}}\xspace}
\providecommand{\mt}{\ensuremath{m_\mathrm{T}}\xspace}

\cmsNoteHeader{EXO-12-057}

\title{Search for heavy Majorana neutrinos in $\mu^\pm \mu^\pm$~+~jets events in proton-proton collisions
at $\sqrt{s} = 8$\TeV}

\date{\today}

\abstract{
A search is performed for heavy Majorana neutrinos (N) using an
event signature defined by two muons of the same charge
and two jets ($\mu^\pm \mu^\pm \mathrm{j j}$). The data correspond to an integrated luminosity of 19.7\fbinv
of proton-proton collisions at a
center-of-mass energy of 8\TeV, collected with the CMS detector at the CERN LHC. No excess of
events is observed beyond the expected standard model background and upper limits are set on
$\abs{\VmN}^2$ as a function of Majorana neutrino mass $\mN$ for masses in the range of
40--500\GeV,
where $\VmN$ is the mixing element of the heavy neutrino with the standard model muon neutrino.
The limits obtained are
$\abs{\VmN}^2 < 0.00470$ for $\mN = 90$\GeV,
$\abs{\VmN}^2 < 0.0123$ for $\mN = 200$\GeV, and
$\abs{\VmN}^2 < 0.583$ for $\mN = 500$\GeV.
These results extend considerably the regions excluded by previous direct searches.
}

\hypersetup{%
pdfauthor={CMS Collaboration},%
pdftitle={Search for heavy Majorana neutrinos in mu+/- mu+/- + jets events in proton-proton collisions
at sqrt(s) = 8 TeV},%
pdfsubject={CMS},%
pdfkeywords={CMS, physics, heavy neutrino}}

\maketitle

\section{Introduction}
The non-zero masses of neutrinos and mixing between flavors have been well established by neutrino oscillation experiments
and provide evidence for physics beyond the standard model (SM)~\cite{pdg}. The most compelling way
of explaining the smallness of the neutrino masses is the ``seesaw" mechanism, which can
be realized in several different schemes~\cite{seesaw1, seesaw2, seesaw3,seesaw4,seesaw5,seesaw6,seesaw7,seesaw8,seesaw9,seesaw10}.
In the simplest model, the SM neutrino
mass is given by $\ m_\nu \approx y^2_\nu v^2/ \mN$, where $y_\nu$ is a Yukawa
coupling of $\nu$ to the Higgs field, $v$ is the Higgs vacuum expectation value in the SM, and
$\mN$ is the mass of a new heavy  neutrino state (N). In this scheme, N is a Majorana particle (which is
its own antiparticle), so processes
that  violate lepton number conservation by two units are possible.

In this paper we describe a search for heavy Majorana neutrinos using a 
phenomenological
approach~\cite{Maj_hadCol_1, Maj_hadCol_2, Maj_hadCol_3, Maj_hadCol_4, Maj_hadCol_5, TaoHan1,Aguila,TaoHan2}.
We follow the studies in Refs. \cite{TaoHan1,Aguila,TaoHan2} and consider a heavy neutrino that
mixes with the SM muon neutrino, with
$\mN$ and $\VmN$ as free parameters of the model. Here
$\VmN$ is a mixing element describing the mixing between the heavy Majorana neutrino and the SM muon neutrino.

Early direct searches for heavy Majorana neutrinos based on this model were reported
by the L3~\cite{l3} and DELPHI~\cite{delphi} experiments at LEP.
They searched for $\cPZ \to \nu_\ell \N$ decays, where $\nu_\ell$ is any SM neutrino ($\ell = \Pe, \mu$, or $\tau$), 
from which limits on $\abs{\VmN}^2$  as a function of $\mN$ can be derived for Majorana neutrino masses up to approximately 90 GeV.
More recently, the CMS experiment at the CERN LHC extended the search region
up to about 200\GeV~\cite{CMS_NR2011}.
Several experiments have obtained limits in the low mass region ($\mN < 5$\GeV), including LHCb~\cite{LHCb_NR} at the LHC.
The searches by L3, DELPHI, and LHCb allow for a finite heavy-neutrino lifetime such that it decays with a vertex 
displaced from the interaction point, while in the search reported here it is assumed that the N decays 
with no significant displacement of the vertex.
Precision electroweak measurements can be
used to constrain the mixing elements, resulting in indirect 90\% confidence level limits of
$\abs{\VmN}^2 < 0.0032$ independent of heavy neutrino mass~\cite{EW_limits}.
Other models with heavy neutrinos have also been examined. ATLAS and CMS at the LHC have reported limits on heavy Majorana
neutrino production~\cite{CMS_LR, ATLAS_NR} in the context of the Left-Right Symmetric Model.
ATLAS have also set limits based on an effective Lagrangian approach~\cite{ATLAS_NR}.

\begin{figure*}[htbp]
  \begin{center}
  \includegraphics[width=0.5\textwidth]{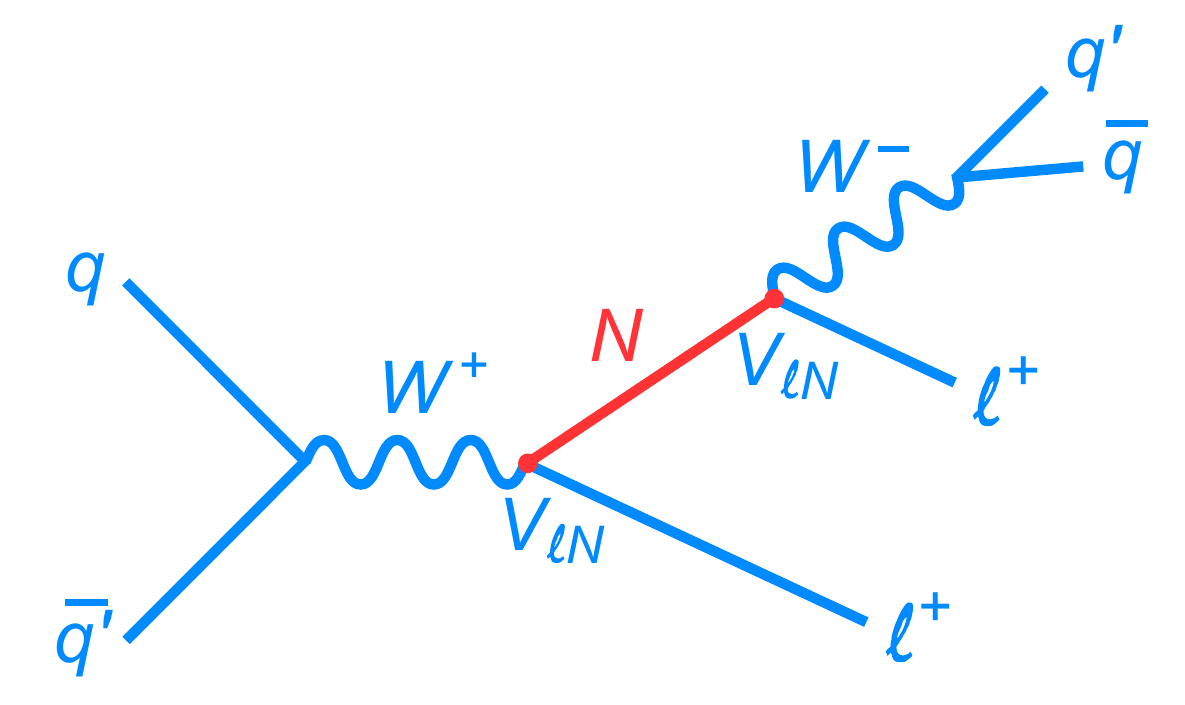}
  \caption{The lowest order Feynman diagram for production of a heavy Majorana neutrino N. 
  The charge-conjugate diagram also contributes and results in a $\ell^- \ell^- \PQq\PAQq'$ final state.  
  }
  \label{fig:feynman}
  \end{center}
\end{figure*}

We report on an updated search for the production of a heavy Majorana neutrino in
proton-proton (pp) collisions at a center-of-mass energy of $\sqrt{s} = 8$\TeV at the LHC with
a data set corresponding to an integrated luminosity of 19.7\fbinv collected with the
CMS detector. We assume that the heavy Majorana neutrino is produced by
$s$-channel production of a W boson, which decays via $\PWp \to \N \Pgmp$. 
The N is assumed to be a
Majorana particle, so it can decay
via $\N \to \PWm \Pgmp$ with $\PWm \to \PQq \PAQq'$, resulting in a
$\Pgmp \Pgmp \PQq \PAQq^\prime$ final state. The principal Feynman diagram for this process is shown in Fig.~\ref{fig:feynman}.
The charge-conjugate decay chain also
contributes and results in a $\Pgmm \Pgmm \PAQq \PQq'$ final state.
The lowest order cross section for $\Pp\Pp \to (\PW^\pm)^* \to \N \mu^\pm$  at a center of mass energy $\sqrt{s}$ is given by
\begin{equation}
\sigma(s) =  \int \rd{}x \int \rd{}y \sum\limits_{q,{\bar q^\prime}}  \left[ f^p_q (x, Q^2) f^p_{\bar q^\prime} (y, Q^2) \right] 
	\hat \sigma  (\hat s)
\end{equation}
where $f^p_q$ are the parton distribution functions for quark $q$ at $Q^2 = \hat s = xys$, and $x$ and $y$ are the fractions of 
the proton momentum carried by the interacting quarks. The parton subprocess cross section $\hat \sigma (\hat s)$ 
is given by~\cite{Pilaftsis}:
\begin{equation}
	\hat \sigma( \hat s) = \frac{\pi \alpha_W^2}{72 {\hat s}^2 (\hat s - m_W^2)^2} \abs{V_{\mu N}}^2 (\hat s - m_N^2)^2 (2 \hat s + m_N^2)
\end{equation} 
where $\alpha_W$ is the weak coupling constant and $m_W$ is the W-boson mass.

We search for events  with two isolated muons of the same sign of electric charge and at least two accompanying jets.
The backgrounds to such dimuon final states originate from SM processes that contain isolated prompt same-sign (SS) dimuons
in the final state (e.g. WZ production), and from
processes such as multijet production in which muons from b-quark decays, or
from jets, are misidentified as isolated prompt muons. The misidentified muon background is more
significant for low masses ($\mN < 90\GeV$), while at higher masses the SM prompt dimuon background becomes more important.

\section{Detector, signal simulation, and data selection}
The CMS detector is described in detail in Ref.~\cite{CMS_detector}.
Its central feature is a superconducting solenoid, which provides a magnetic field of 3.8\unit{T} along
the direction of the counterclockwise rotating beam, taken as the $z$~axis of the detector coordinate system,
with the center of the detector defined to be at $z = 0$. The azimuthal angle $\phi$ is measured in the
plane perpendicular to the $z$ axis, while the polar angle $\theta$ is measured with respect to this axis.
Within the superconducting solenoid volume are a silicon pixel and
strip tracker, a lead tungstate crystal electromagnetic calorimeter, and a brass and
scintillator hadron calorimeter, each composed of a barrel and two endcap sections. Muons are measured in
gas-ionization detectors embedded in the steel flux-return yoke outside the solenoid. Extensive
forward calorimetry complements the coverage provided by the barrel and endcap detectors.
A two-level trigger system selects the most interesting events for analysis.

Muons are measured in the pseudorapidity range $\abs{\eta}< 2.4$, defined as $\eta = - \ln [\tan (\theta / 2)]$,
with detection planes made using three technologies:
drift tubes, cathode strip chambers, and resistive plate chambers.
Charged-particle trajectories are measured in a silicon pixel and strip tracker
covering $0 \le \phi \le 2 \pi$ in azimuth and $\abs{\eta} < 2.5$.
Matching muons to tracks measured in the silicon tracker
results in a relative transverse momentum resolution for muons with $20 < \pt < 100\GeV$ of 1.3--2.0\% in the barrel and
better than 6\% in the endcaps. The $\pt$ resolution in the barrel is better than 10\% for muons
with $\pt$ up to 1\TeV~\cite{mu_reco}.

The heavy Majorana neutrino production and decay process are simulated using the leading-order (LO) event generator described in
Ref.~\cite{Aguila} and implemented in \ALPGEN~v2.14~\cite{ALPGEN}.
The LO cross section at $\sqrt{s} = 8$\TeV for $\Pp\Pp \to \N \mu^\pm \to \mu^\pm \mu^\pm \PQq \PAQq^\prime$
with $\abs{\VmN}^2 = 1$ has a value of
1515\unit{pb} for $\mN = 40$\GeV, dropping to
3.56\unit{pb} for $\mN = 100$\GeV, and to
2.15\unit{pb} for $\mN = 500$\GeV~\cite{Aguila}.
The cross section is proportional to $\abs{\VmN}^2$.
We scale the leading order cross section by a factor of 1.34 to account for higher-order corrections, based
on the next-to-next-to-leading-order calculation in Refs.~\cite{FEWZ1, FEWZ2} for $s$-channel $\PWpr$
production that has the same production kinematics as the signal.
We use the CTEQ6M parton distribution functions (PDFs)~\cite{CTEQ6}.
Parton showering and hadronization are simulated using \PYTHIA~v6.4.22~\cite{pythia}.  The Monte Carlo (MC)
event generator is interfaced with CMS software, where {\GEANTfour}~\cite{Geant4} detector simulation,
digitization of simulated electronic signals, and event reconstruction are performed.
To ensure correct simulation of the number of additional interactions per bunch crossing (pileup),
Monte Carlo simulated events are mixed with multiple minimum bias events with weights chosen
using the distribution of the number of reconstructed pp interaction vertices observed in data.

A dimuon trigger is used to select the signal sample. This trigger requires the presence of one muon
reconstructed from matching tracks in the muon system and tracker with
transverse momentum ($\pt$) above 17\GeV
and a second muon with $\pt > 8$\GeV formed from a track in the tracker matched to hits in one of the muon detectors.
The trigger efficiency for the signal sample is measured using $\cPZ\to \Pgmp \Pgmm$ events selected in data,
and is found to be in the range 85--95\%, varying with $\pt$ and $\eta$.

Additional selections are performed to ensure the presence of well-identified muons and jets. Events
are first required to have a well-reconstructed pp interaction vertex (primary vertex)
identified as the reconstructed vertex with the largest value of $\sum \pt^2$ for its
associated charged tracks reconstructed in the tracking detectors~\cite{CMS_tracker}.

Leptons, jets, and missing transverse energy (\ETslash)  in the event are reconstructed using the standard
CMS particle-flow techniques~\cite{pflow1, pflow2}.
The missing transverse momentum vector is defined as the projection on the plane perpendicular to the beams of
the negative vector sum of the momenta of all reconstructed particles in an event. Its magnitude is referred to as \ETslash.
Jets are formed from clusters based on the anti-\kt algorithm~\cite{antiKT}, with a distance
parameter of 0.5, and are required to be within the pseudorapidity range $\abs{\eta} < 2.5$.
Muon candidates are required to have $\abs{\eta} < 2.4$ and to be consistent with
originating from the primary vertex. Muon candidates are reconstructed by matching tracks in
the silicon tracker to hits in the outer muon system, and are also required to satisfy specific track quality
and calorimeter deposition requirements~\cite{mu_reco}.
Muon candidates must be isolated from other activity in the event, which is insured by requiring their
relative isolation parameter  ($I_\text{rel}$) to be less than 0.05. Here $I_\text{rel}$ is defined as the scalar sum of transverse
energy present within
$\Delta R = \sqrt{\smash[b]{(\Delta \eta)^2 + (\Delta \phi)^2}} < 0.3$ of the candidate's
direction, excluding the candidate itself, divided by the muon candidate's transverse momentum. The muon selection criteria
are the same as those used in Ref.~\cite{SSsusy2014} except for the more stringent requirement on $I_\text{rel}$ used here.

Events are required to contain two same-sign muons, one  with $\pt > 20$\GeV
and the other with $\pt > 15$\GeV, and at least two jets with $\pt > 20$\GeV.
Events with a third muon are rejected to suppress diboson events such as WZ.
To reduce backgrounds from top-quark decays, events in which at least one jet is identified as originating
from a b quark are rejected, where the medium working point of the combined secondary vertex tagger~\cite{bTag} has been used.

Further  selection requirements are made based on two Majorana neutrino mass search regions. In the low-mass
search region
($\mN \lesssim 80\GeV$) the W-boson propagator is on-shell and the final state system
of $\mu^\pm \mu^\pm \PQq \PAQq^\prime$ should have an invariant mass close to $m_\PW$.
In this region the following
selections are imposed: missing transverse energy $<$30\GeV;
$\mu^\pm \mu^\pm \mathrm{j j}$ invariant mass $<$200\GeV, where the two jets chosen are those
which result in $m(\mu^\pm \mu^\pm \mathrm{j j})$ closest to $m_\PW$; and dijet invariant mass of these two jets
$m(\mathrm{jj}) < 120\GeV$.
In the low-mass region the transverse momenta of the muons and jets are relatively low and the
overall efficiency is considerably smaller than at higher masses. Therefore,  the invariant mass requirements in this region
are chosen to be relatively loose.

In the high-mass search region ($\mN \gtrsim 80\GeV$) the W-boson propagator is off-shell but the W boson from the N decay is
on-shell, so the $\PW \to \PQq \PAQq^\prime$ decay should result in a dijet invariant mass close to $m_\PW$.
As the decay particles are more energetic in this region, the following selections are imposed: missing transverse
energy less than 35\GeV (relaxed to increase signal efficiency);
dijet invariant mass $50 < m(\mathrm{jj}) < 110\GeV$, where the two jets with the invariant mass
closest to $m_\PW$ are chosen. In both the low-mass and high-mass search regions,
the upper cutoff on the missing transverse energy suppresses SM background processes in which a W boson decays leptonically
($\PW \to \mu \nu$), including W~+~jet and $\ttbar$ production.

The low-mass selection gives the best sensitivity for masses below 90\GeV, while for $\mN \ge 90\GeV$ the
high-mass selection gives the best sensitivity. Therefore, we use the low-mass selection for $40 < \mN < 90\GeV$  and the
high-mass selection for $\mN \ge 90\GeV$. For masses less than 40\GeV the overall acceptance for the Majorana neutrino signal
is less than 1\% and we do not search in this region.

After applying all the selection criteria above, a final selection is applied based on optimizing the
signal significance using a figure of merit~\cite{punzi} defined by $\epsilon_S / (a/2 + \delta B)$
with the number of standard deviations $a=2$
and where $\epsilon_S$ is the signal selection efficiency and $\delta B$ is the uncertainty in the estimated background.

At each Majorana neutrino mass point,
the optimization is performed  by varying the lower bound of three selections:
the transverse momentum $p_{\mathrm{T}1}$ of the highest $\pt$ muon (or ``leading" muon), the
transverse momentum $p_{\mathrm{T}2}$ of the second muon (or ``trailing" muon), and the invariant mass $m(\mu^\pm \mu^\pm \mathrm{j j})$.
The two jets used in this optimization are those selected as described above for the low- and high-mass regions.

The overall signal acceptance includes trigger efficiency, geometrical acceptance, and efficiencies of all selection criteria.
The overall acceptance for heavy Majorana neutrino events ranges between 0.69\% for $\mN = 40$\GeV
to 12\% for $\mN = 500$\GeV. The lower acceptance at low $\mN$ is due to the
smaller average $\pt$ of the jets and muons in these events. The acceptance dips when $\mN$ approaches $m_\PW$ because
the muon $\pt$ becomes small. Above the W-boson mass the acceptance increases again up to 400--500\GeV
at which point the boosted decay products of the Majorana neutrino begin to overlap.

\section{Background estimation}
There are three potential sources of same-sign dimuon backgrounds:  SM sources such as WZ production,
events resulting from misidentified muons,
and opposite-sign dimuon events (e.g. from $\cPZ\to \Pgmp \Pgmm$) in which the charge of
one of the muons is mismeasured. The latter source is found to be negligible for muons with $\pt$ in the
range of interest for this analysis, based on MC
studies and studies with cosmic ray muons. Estimation of the remaining two sources of background is discussed below.

An important background source is the irreducible background from SM production of two genuine
isolated muons of the same sign, which can originate from sources such as
WZ and ZZ diboson production,
double W-strahlung $\PW^\pm \PW^\pm \PQq \PQq$ processes,
WH production, $\ttbar\PW$ production,
double parton scattering (two $\PQq \PQq^\prime \to \PW)$,
and triboson production.
These processes have relatively small cross sections,
and are consequently estimated using MC simulations. We use \PYTHIA to simulate ZZ and WZ production and
\MADGRAPH~v5.1.3.30~\cite{madgraph} for the remaining processes.

The second significant background source originates from
events containing objects misidentified as prompt muons. These ``prompt muons''
originate from b-quark decays or light-quark or gluon jets.
Examples of this background include: multijet production in which two jets are misidentified as muons;
$\PW(\to \mu \nu)$~+~jets events in which one of the jets is misidentified as a muon; and
$\ttbar$ decays in which one of the top-quark decays yields a prompt isolated muon
($\mathrm{t} \to \PW \mathrm{b} \to \mu \nu_\mu \mathrm{b}$), and the other muon of same charge
arises from a b-quark decay or a jet misidentified as an isolated prompt muon.
These backgrounds are estimated using
control samples from collision data as described below.

To estimate the misidentified muon background we use the method described in Ref.~\cite{CMS_NR2011}.
An independent data sample enriched in multijet events is used to
calculate the probability for a jet that passes minimal muon selection requirements (``loose muons'') to also pass
the more stringent requirements used to define muons selected in the heavy Majorana neutrino signal selection (``tight muons'').
This probability is binned in $\pt$ and $\eta$ and is used as a weight in the calculation of the background in
events that pass all the signal selections except that one or both muons
fail the tight criteria, but pass the loose ones.

The sample enriched in multijet production is selected by requiring a loose muon and a jet, resulting in events that are mostly dijet events with one jet
containing a muon. Only one muon is allowed and  upper cutoffs on missing transverse energy ($\ETslash < 20$\GeV)
and the transverse mass ($\mt < 25$\GeV) are applied, where $\mt$ is calculated using the muon $\pt$ and $\ETslash$.
These requirements suppress contamination from $W$ and $Z$ boson decays. We also require the loose
muon and jet to be separated in azimuth by $\Delta \phi > 2.5$. This away-side jet is used as a tag and the loose muon
is a probe used to determine the misidentification probability. The transverse energy of the tag jet is
an essential ingredient to calibrate the characteristics of the probe (loose muon).

Loose muons are defined by relaxing the identification requirements (used to select signal events)  as follows:
the isolation requirement is relaxed from $I_\text{rel} < 0.05$ to $I_\text{rel} <0.4$;
the transverse impact parameter of the muon track is relaxed from $<$0.05\unit{mm} to $<$2\unit{mm};
the chi squared per degree of freedom of the muon track fit is relaxed from 10 to 50.

The overall systematic uncertainty in the misidentified muon background is determined from
the variation of the background estimate with respect to the isolation requirement for
the loose muons and the $\pt$ requirement for the tagging jet.
Increasing and decreasing the $\pt$ requirement for the tag jet changes the $\pt$ spectrum
of the recoiling muon in the event and is found to have the largest impact on the
background level.  The misidentified muon rate is comparatively stable with respect to
variations in loose muon isolation requirements.  As a result, the 28\% overall systematic uncertainty
in the misidentified muon background estimate is dominated by the $\pt$ requirement on the tag jet.

We evaluate the method used to estimate the background from misidentified muons by checking the
procedure using MC simulated event samples in which the true origin of the muons,
either from W or Z boson decays  or from a quark decay, is known.
The misidentification probabilities are obtained from multijet events and are used to estimate the
misidentified muon backgrounds in $\ttbar$, W~+~jets, and independent multijet events by applying the
background estimation method described above.  The predicted backgrounds agree with the
expectations within the systematic uncertainties.

In addition, to test the validity of the background estimation method, we define
two signal-free control regions in data. We apply the background estimation method in these regions and
compare the result with the observed yields. The control regions in the two mass ranges are defined as follows.
In both the low-mass range ($40 < \mN < 90\GeV$) and the high-mass range ($\mN > 90\GeV$)
the control region is the same as the signal selection, without the final optimized selections
but with either missing transverse
energy greater than 50\GeV or one or more jets that are tagged as originating from a b quark.
In the low-mass control region we predict  a total background of
$51.4 \pm 1.9\stat \pm 8.3\syst$
events compared to an observed yield of 45 events, while in the high mass control region we predict a
total background of $87.6 \pm 2.5\stat \pm 12.3\syst$
compared to the observed yield of 81 events.
The misidentified muon background accounts for about $2/3$ of the total background in both regions.
The remaining backgrounds originate mainly from diboson production (approximately 25\% of the total background)
and Higgs boson production (approximately 4\%).
In both regions the predictions are in agreement with the observations
and well within the systematic uncertainty, which is dominated by the
28\% uncertainty in the misidentified muon background.
The observed distributions of all relevant observables also agree with the predictions, within uncertainties.

\section{Systematic uncertainties}
The main sources of systematic uncertainties are associated with the background estimates. As described above,
the overall systematic uncertainty in the misidentified muon background is 28\%.
The systematic uncertainties in the normalizations of irreducible SM backgrounds are 12\% for
WZ and 9\% for ZZ~\cite{CMS_ZZ}. 
For the other processes the uncertainty is 25\%,
determined by varying the renormalization and factorization scales from the nominal value of $Q^2$
to $4Q^2$ and $Q^2/4$, and following the PDF4LHC recommendations~\cite{pdf1, pdf2} to estimate the uncertainty due to
the choice of PDFs.
After combining all the SM backgrounds with their respective uncertainties, the overall systematic uncertainty 
in the SM irreducible background estimate, including those discussed below, is 19\%.

Sources of systematic uncertainty associated with the estimates of the heavy Majorana neutrino signal and of the SM irreducible
background are summarized in Table~\ref{table:syst}.
\begin{table*}[tbh]
  \topcaption{Summary of systematic uncertainties in the estimation of the heavy Majorana signal,
  SM background (SM Bkgd.), and misidentified muon background (Misid. Bkgd.) for low-mass selection (first number) and
  high mass selection (second number in square brackets).}
  \label{table:syst}
  \centering
  \begin{tabular}{llll}
  \hline
  Source & Signal (\%)  & SM Bkgd. (\%)  & Misid. Bkgd. (\%) \\
  \hline
Background estimate from data	& --- 			& --- 				& 28 [28] \\
SM cross section  			& ---					& 9--25 [9--25] 	& --- \\
Jet energy scale 			& 5--12 [1--3] 	& 3 [5] 			& ---	\\
Jet energy resolution  		& 6--12 [2--3]	& 7 [10] 			& ---	\\
Event pileup 		   		& 6--7 [3--6]		& 7 [4] 			& ---	\\
Unclustered energy			& 2 [1]				& 3 [1]				& --- \\
Integrated luminosity 		& 2.6 [2.6]			& 2.6 [2.6] 		& ---	\\
Muon trigger and selection 	& 2 [2] 			& 2 [2] 			& ---	\\
b-tagging 						& 1 [1] 			& 1.5 [1] 			& --- \\
PDF							& 3.5 [3.5]			& ---					& --- \\
Renormalization/Factorization scales					& 8--10 [1--6]		& --- 				& --- \\
Signal MC statistics 		& 3--9 [1--4] 		& --- 				& --- \\

  \hline
\end{tabular}
\end{table*}
To evaluate the uncertainty due to imperfect knowledge of the integrated luminosity~\cite{lumi},
jet energy scale~\cite{jes}, jet energy resolution~\cite{jes},
b-tagging~\cite{bTag}, muon trigger and selection efficiency, and the cross section for minimum bias production used in the
pileup reweighting procedure in simulation, the input value of each parameter is changed by $\pm$1 standard deviation
from its central value.
Energy not clustered in the detector affects the overall missing transverse energy scale resulting in an
uncertainty in the event yield due to the upper cutoff on $\ETslash$.
Additional uncertainties in the heavy Majorana neutrino signal estimate arise from the choice of PDFs
and renormalization and factorization scales used in the \ALPGEN MC event generator.

\section{Results and discussion}

\begin{figure}[tbph]
\centering
	\includegraphics[width=\cmsFigWidthThree]{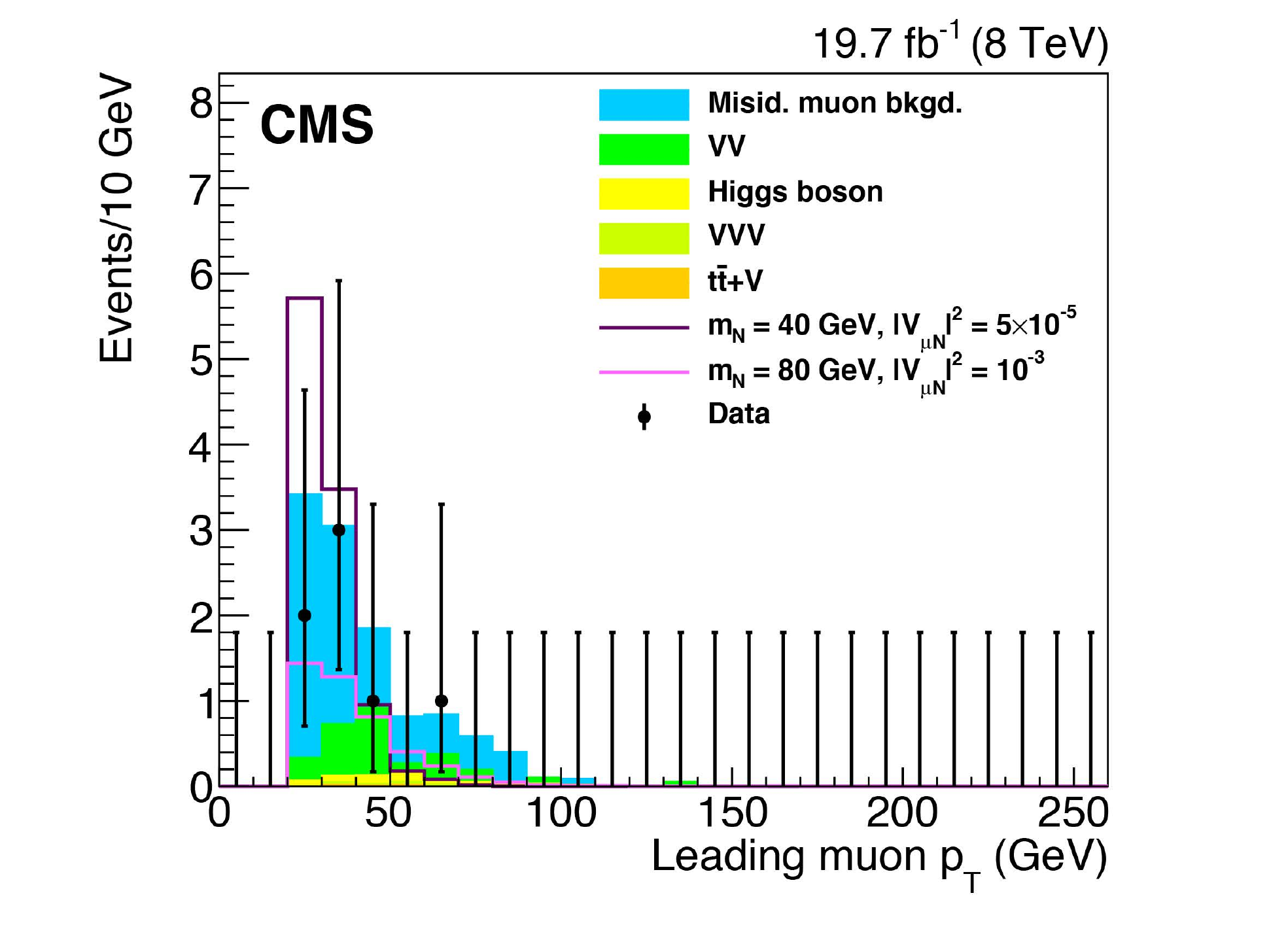}
	\includegraphics[width=\cmsFigWidthThree]{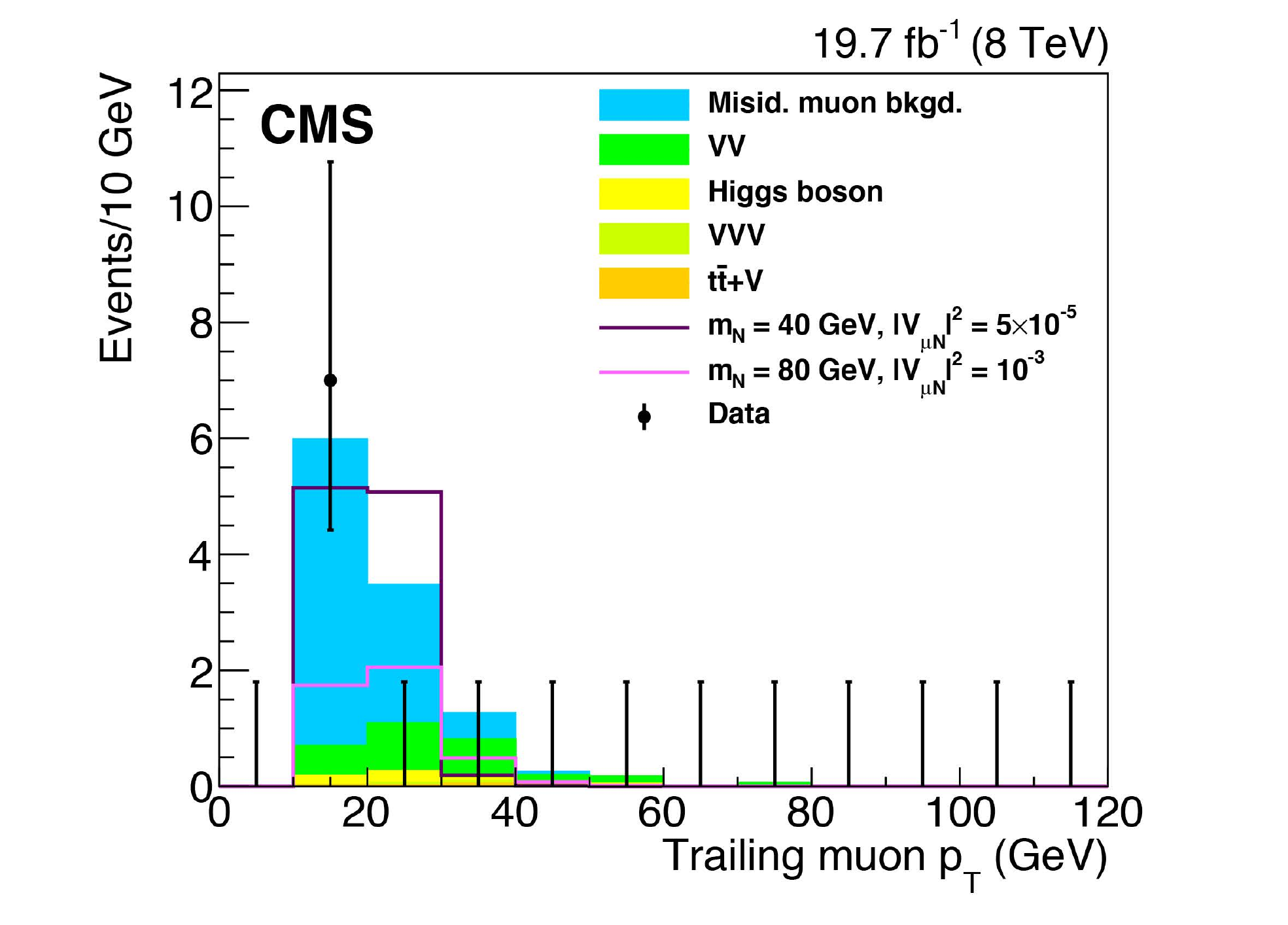}
	\includegraphics[width=\cmsFigWidthThree]{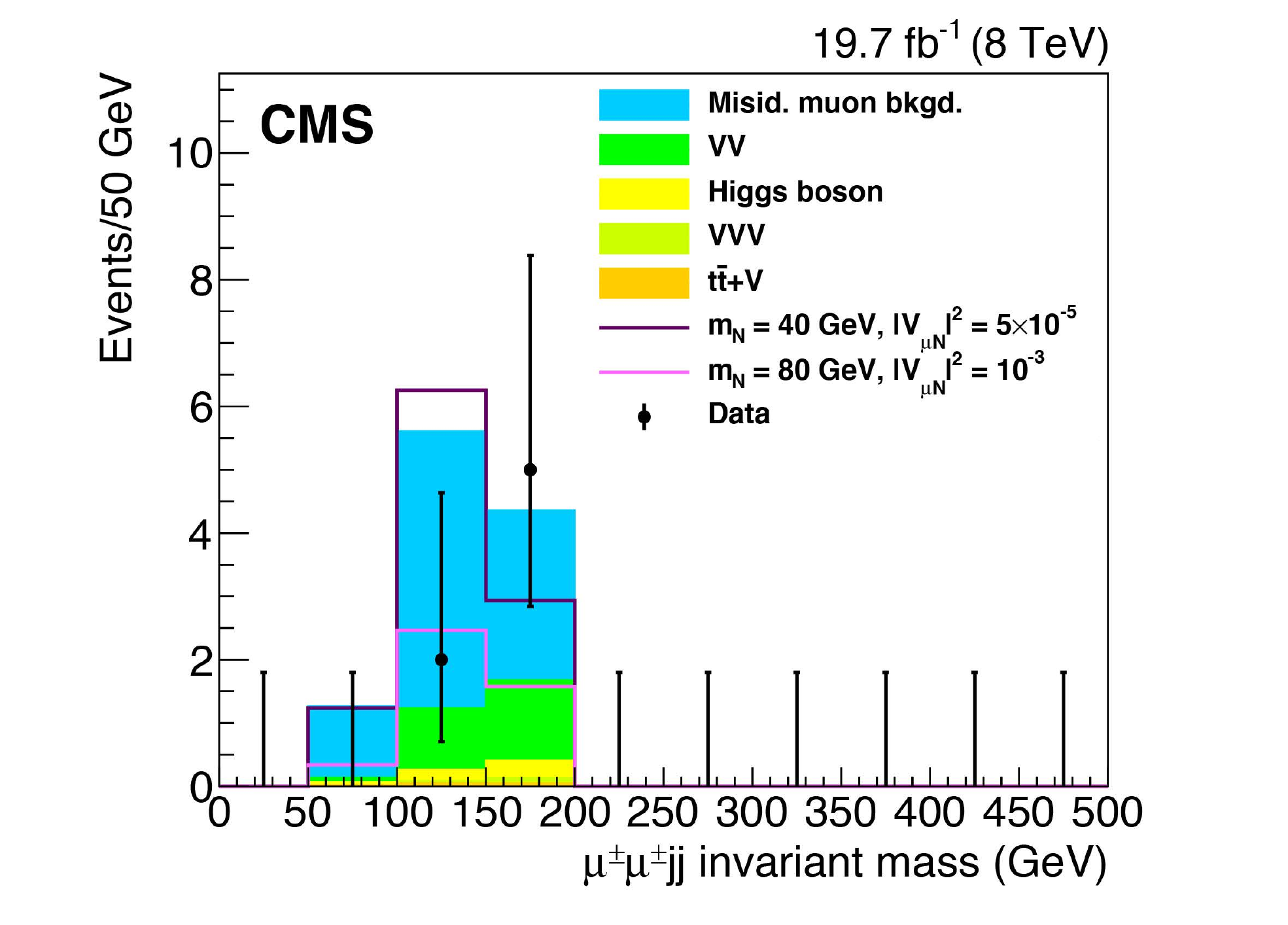}
	\caption{Kinematic distributions for the low-mass region after all selections are applied
	except for the final optimization requirements:
	leading muon $\pt$ (top),  trailing muon $\pt$ (middle), and
	$\mu^\pm \mu^\pm \mathrm{j j}$ invariant mass (bottom).
	The plots show the data, backgrounds, and two choices for the
	heavy Majorana neutrino signal:  $\mN = 40$\GeV, $\abs{\VmN}^2 = 5\times 10^{-5}$ and
	$\mN = 80$\GeV, $\abs{\VmN}^2 = 1 \times 10^{-3}$.
	The backgrounds shown are from misidentified muons and from diboson (VV), Higgs boson, triboson (VVV),
	and $\ttbar\PW$ production.
	}
	\label{fig:lowmass_plots}

\end{figure}

\begin{figure}[tbh*]
\centering
	\includegraphics[width=\cmsFigWidthThree]{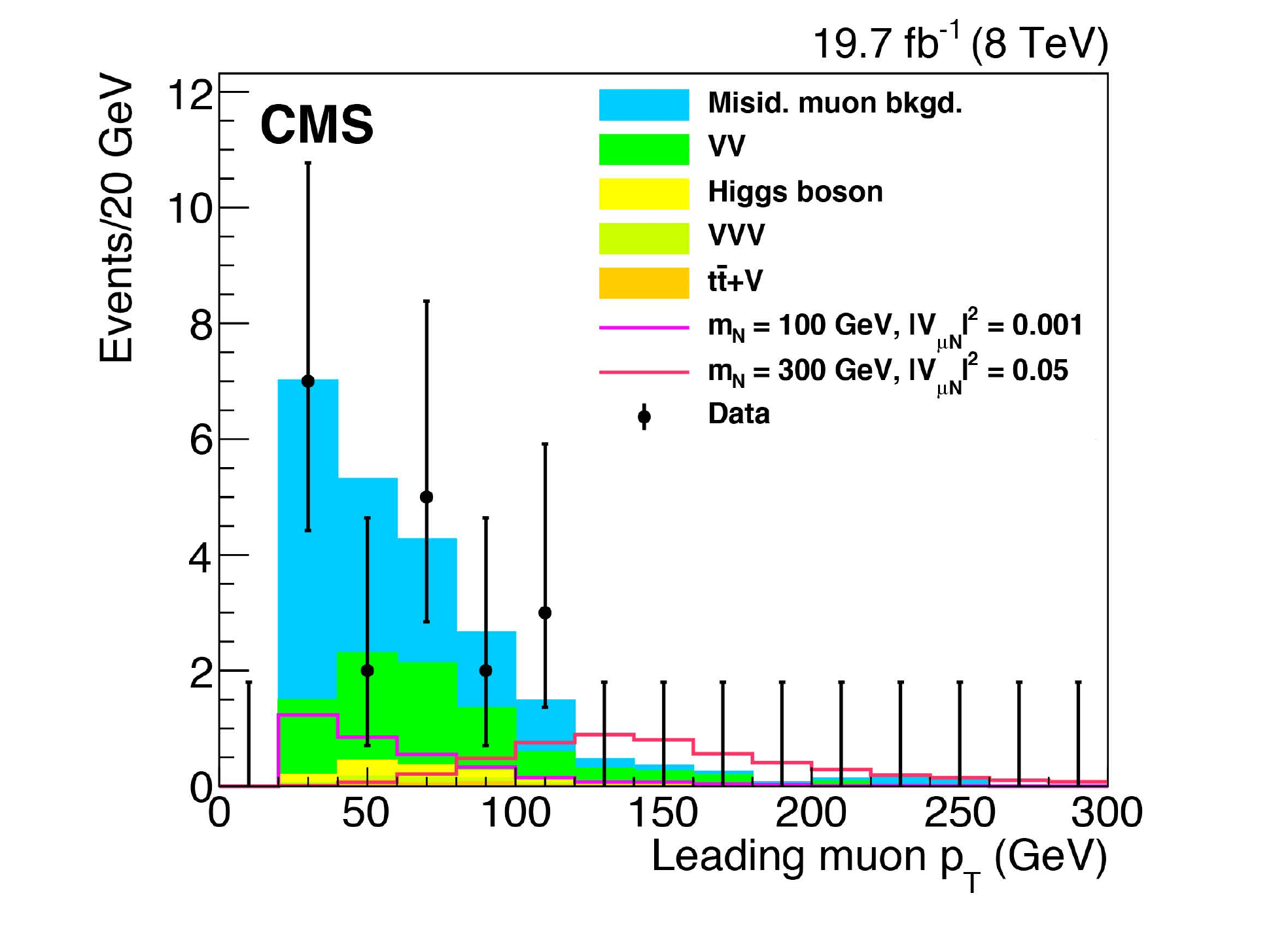}
	\includegraphics[width=\cmsFigWidthThree]{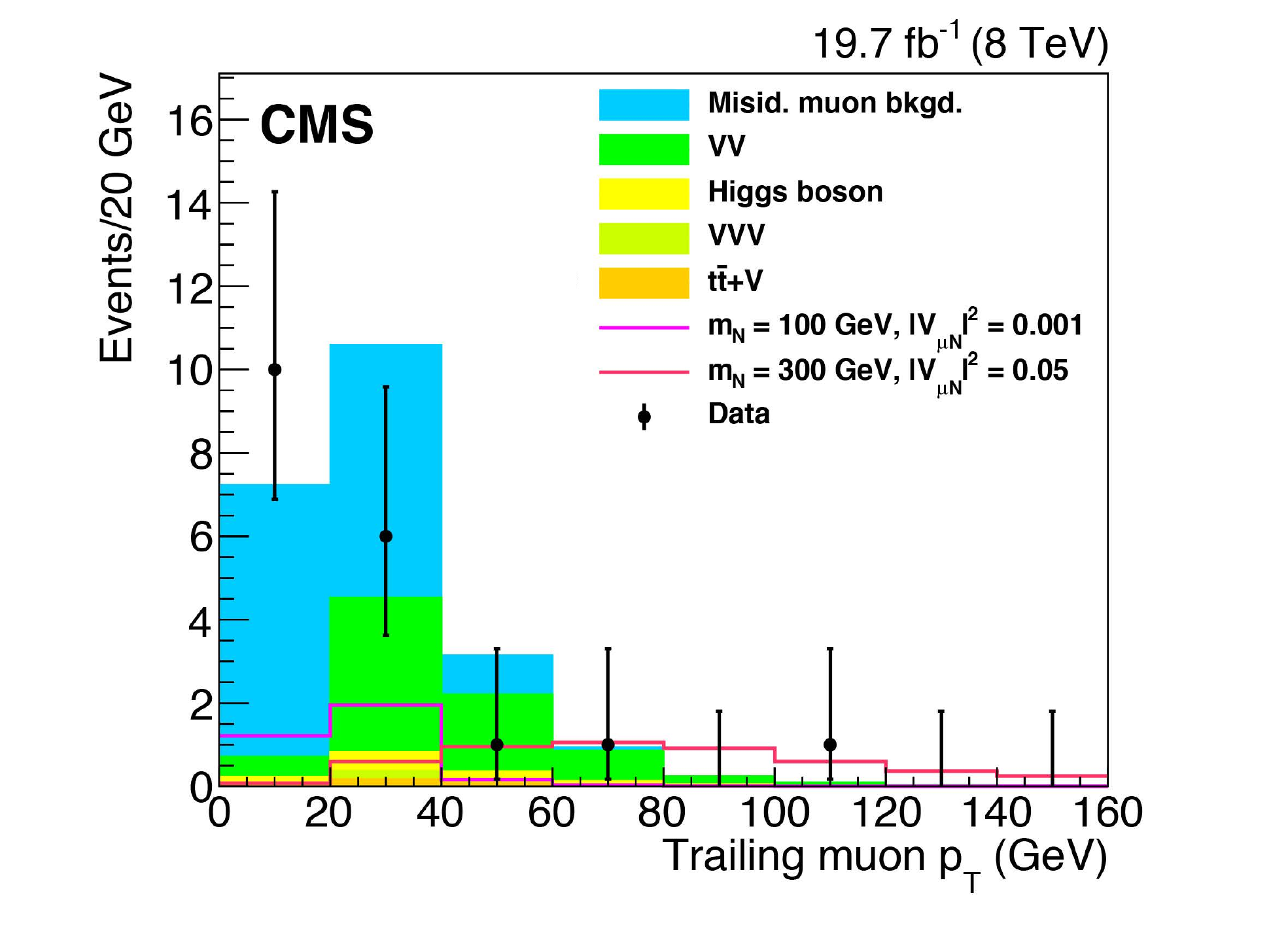}
	\includegraphics[width=\cmsFigWidthThree]{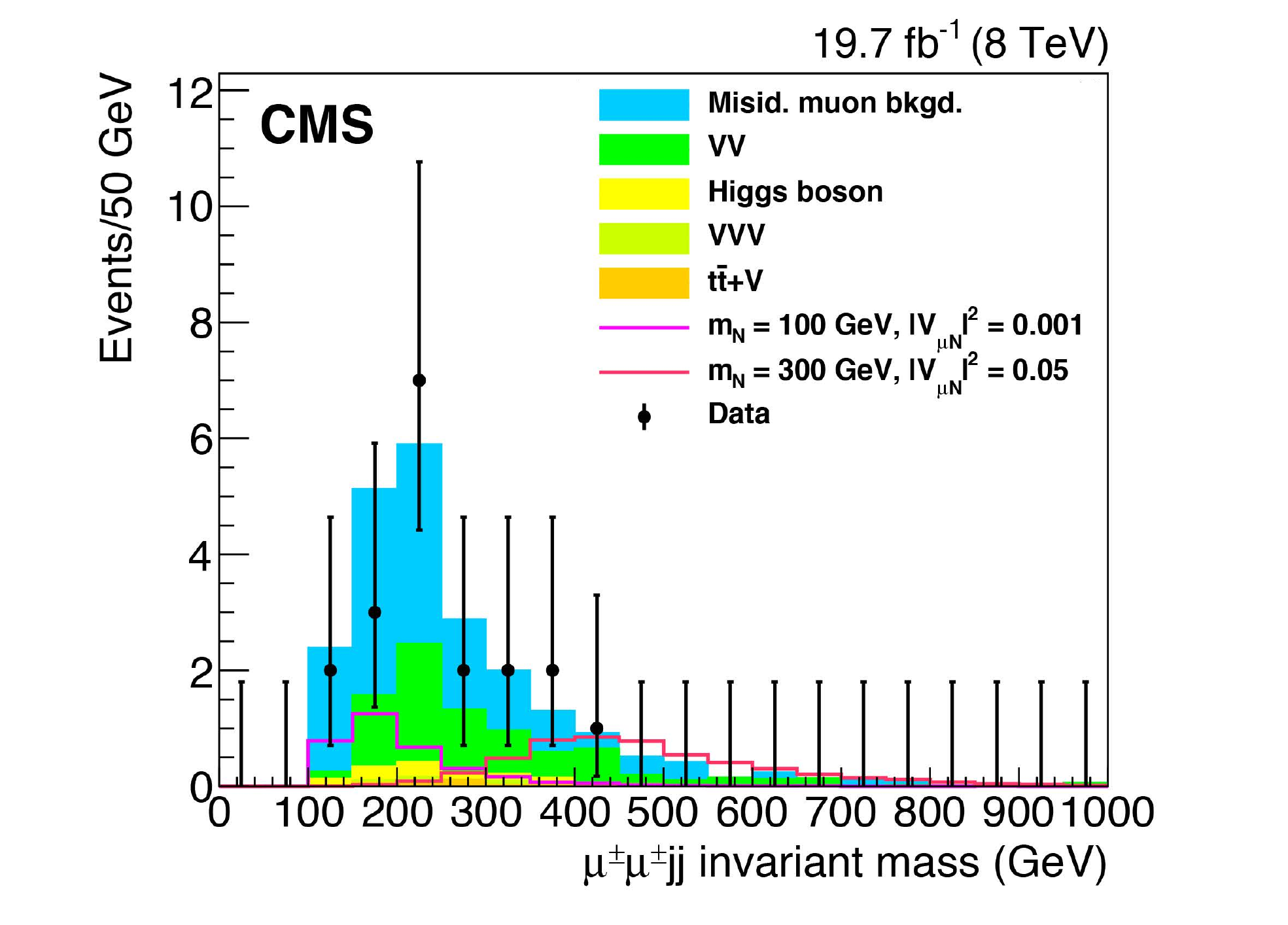}
	\caption{Kinematic distributions for the high-mass region after all selections are applied except
	for the final optimization requirements:
	leading muon $\pt$ (top),  trailing muon $\pt$ (middle), and
	$\mu^\pm \mu^\pm \mathrm{j j}$ invariant mass (bottom).
	The plots show the data, backgrounds, and two choices for the
	heavy Majorana neutrino signal:  $\mN = 100$\GeV, $\abs{\VmN}^2 = 0.001$ and
	$\mN = 300$\GeV, $\abs{\VmN}^2 = 0.05$.
	The backgrounds shown are from misidentified muons and from diboson (VV), Higgs boson, triboson (VVV),
	and $\ttbar \PW$ production.
	}
	\label{fig:highmass_plots}

\end{figure}

The selections determined by the optimization for each Majorana neutrino mass point are shown
in Table~\ref{table:optimize} together with the overall signal acceptance.
\begin{table}[tbh]
  \topcaption{Minimum thresholds on discriminating variables determined by the optimization and overall signal acceptance
  for each Majorana neutrino mass point $(\mN)$. As discussed in the text, the search was divided into low-mass ($\mN \le 80\GeV$) and high-mass ($\mN > 80\GeV$) regions, and different selection criteria were used in the two  regions. }
  \label{table:optimize}
  \centering
  \begin{tabular}{ccccc}
  \hline
  $\mN$ & $m(\mu^\pm \mu^\pm \mathrm{j j})$  & $p_{\mathrm{T}1}$  & $p_{\mathrm{T}2}$ & Acceptance \\
  (\GeVns) & (\GeVns) &  (\GeVns)  & (\GeVns) & (\%)\\
  \hline
40 	& 80 		& 20	& 15	& 0.69 \\
50 	& 80 		& 20 	& 15	& 0.80 \\
60  & 80 		& 20 	& 15	& 0.64 \\
70 	& 80 		& 20 	& 15	& 0.26 \\
80 	& 80		& 20 	& 15	& 1.2 \\
90 		& 110 	& 20 	& 15	& 1.2   \\
100 	& 120 	& 20 	& 15	& 4.7   \\
125 	& 140 	& 25 	& 20	& 11  \\
150 	& 160 	& 35 	& 25	& 13	 \\
175 	& 200 	& 45 	& 30	& 15  \\
200 	& 220 	& 50 	& 35	& 16	 \\
250 	& 270 	& 75 	& 35	& 17	 \\
300 	& 290 	& 100 	& 45	& 15	 \\
350 	& 290 	& 100 	& 45	& 16	 \\
400 	& 290 	& 100 	& 45	& 15	 \\
500 	& 290 	& 100 	& 45	& 12 \\
  \hline
\end{tabular}

\end{table}
Figure~\ref{fig:lowmass_plots} shows the transverse momentum distributions of the two muons, $p_{\mathrm{T}1}$ and $p_{\mathrm{T}2}$,
and the invariant mass $m(\mu^\pm \mu^\pm \mathrm{j j})$ for
the low-mass region after all selections are applied except for the final optimization requirements.
The corresponding distributions for the high-mass region are shown  in Fig.~\ref{fig:highmass_plots}.

After applying all the final optimized selections we obtain the results shown in Table~\ref{table:yields}.
The expected signal depends on $\mN$ and $\abs{\VmN}^2$. For $\mN =50$\GeV and
$\abs{\VmN}^2 = 1 \times 10^{-3}$ the expected number of events is 226.
For $\mN = 100$\GeV and  $\abs{\VmN}^2 = 1 \times 10^{-3}$ the expected number of events
is 4.4, while for $\mN =500$\GeV and $\abs{\VmN}^2 = 1$ it is 6.9.

\begin{table*}[tbh]
  \topcaption{Observed event yields and estimated backgrounds for each Majorana neutrino mass point $(\mN)$.
  The background predictions from the irreducible SM backgrounds (SM Bkgd.),
  misidentified muon background (Misid. Bkgd.) and total background (Total Bkgd.) are shown in columns 2--4,
  while column 5 shows the number of events observed in data. The uncertainties shown are respectively
  the statistical and systematic components. As discussed in the text, the search was divided into low-mass ($\mN \le 80\GeV$) and high-mass ($\mN > 80\GeV$) regions, and different selection criteria were used in the two  regions.}
  \label{table:yields}
  \centering
  \begin{tabular}{ccccc}
  \hline
  $\mN$ & SM Bkgd. & Misid. Bkgd. & Total Bkgd.  & $N_\text{obs}$ \\
  (\GeVns) &  & &  & \\
  \hline
40    & 3.0 $\pm$ 0.4 $\pm$ 0.6  & 6.7 $\pm$ 0.9 $\pm$ 1.9  & 9.8 $\pm$ 1.0 $\pm$ 2.0   & 7 \\
50    & 3.0 $\pm$ 0.4 $\pm$ 0.6  & 6.7 $\pm$ 0.9 $\pm$ 1.9  & 9.8 $\pm$ 1.0 $\pm$ 2.0   & 7 \\
60    & 3.0 $\pm$ 0.4 $\pm$ 0.6  & 6.7 $\pm$ 0.9 $\pm$ 1.9  & 9.8 $\pm$ 1.0 $\pm$ 2.0   & 7 \\
70    & 3.0 $\pm$ 0.4 $\pm$ 0.6  & 6.7 $\pm$ 0.9 $\pm$ 1.9  & 9.8 $\pm$ 1.0 $\pm$ 2.0   & 7 \\
80    & 3.0 $\pm$ 0.4 $\pm$ 0.6  & 6.7 $\pm$ 0.9 $\pm$ 1.9  & 9.8 $\pm$ 1.0 $\pm$ 2.0   & 7 \\
90    & 8.7 $\pm$ 0.7 $\pm$ 1.7  & 12.6 $\pm$ 1.1 $\pm$ 3.5  & 21.3 $\pm$ 1.3 $\pm$ 3.9  & 19 \\
100   & 8.7 $\pm$ 0.7 $\pm$ 1.7  & 11.7 $\pm$ 1.0 $\pm$ 3.3  & 20.4 $\pm$ 1.2 $\pm$ 3.7  & 19 \\
125   & 7.9 $\pm$ 0.6 $\pm$ 1.5  & 5.9 $\pm$ 0.7 $\pm$ 1.6   & 13.8 $\pm$ 0.9 $\pm$ 2.2  & 8 \\
150   & 6.4 $\pm$ 0.5 $\pm$ 1.2  & 3.6 $\pm$ 0.6 $\pm$ 1.0   & 9.9 $\pm$ 0.8 $\pm$ 1.6   & 7 \\
175   & 4.4 $\pm$ 0.4 $\pm$ 0.8  & 1.6 $\pm$ 0.4 $\pm$ 0.5   & 6.0 $\pm$ 0.6 $\pm$ 1.0   & 7 \\
200   & 3.4 $\pm$ 0.4 $\pm$ 0.7  & 0.8 $\pm$ 0.3 $\pm$ 0.2   & 4.2 $\pm$ 0.5 $\pm$ 0.7   & 5 \\
250   & 1.9 $\pm$ 0.3 $\pm$ 0.3  & 0.6 $\pm$ 0.2 $\pm$ 0.2   & 2.5 $\pm$ 0.3 $\pm$ 0.4   & 3 \\
300   & 0.9 $\pm$ 0.2 $\pm$ 0.2  & 0.1 $\pm$ 0.2 $\pm$ 0.0   & 1.0 $\pm$ 0.3 $\pm$ 0.2   & 1 \\
350   & 0.9 $\pm$ 0.2 $\pm$ 0.2  & 0.1 $\pm$ 0.2 $\pm$ 0.0   & 1.0 $\pm$ 0.3 $\pm$ 0.2   & 1 \\
400   & 0.9 $\pm$ 0.2 $\pm$ 0.2  & 0.1 $\pm$ 0.2 $\pm$ 0.0   & 1.0 $\pm$ 0.3 $\pm$ 0.2   & 1 \\
500   & 0.9 $\pm$ 0.2 $\pm$ 0.2  & 0.1 $\pm$ 0.2 $\pm$ 0.0   & 1.0 $\pm$ 0.3 $\pm$ 0.2   & 1 \\
  \hline

\end{tabular}

\end{table*}

\begin{figure}[h*btp]\centering
  \includegraphics[width=\cmsFigWidth]{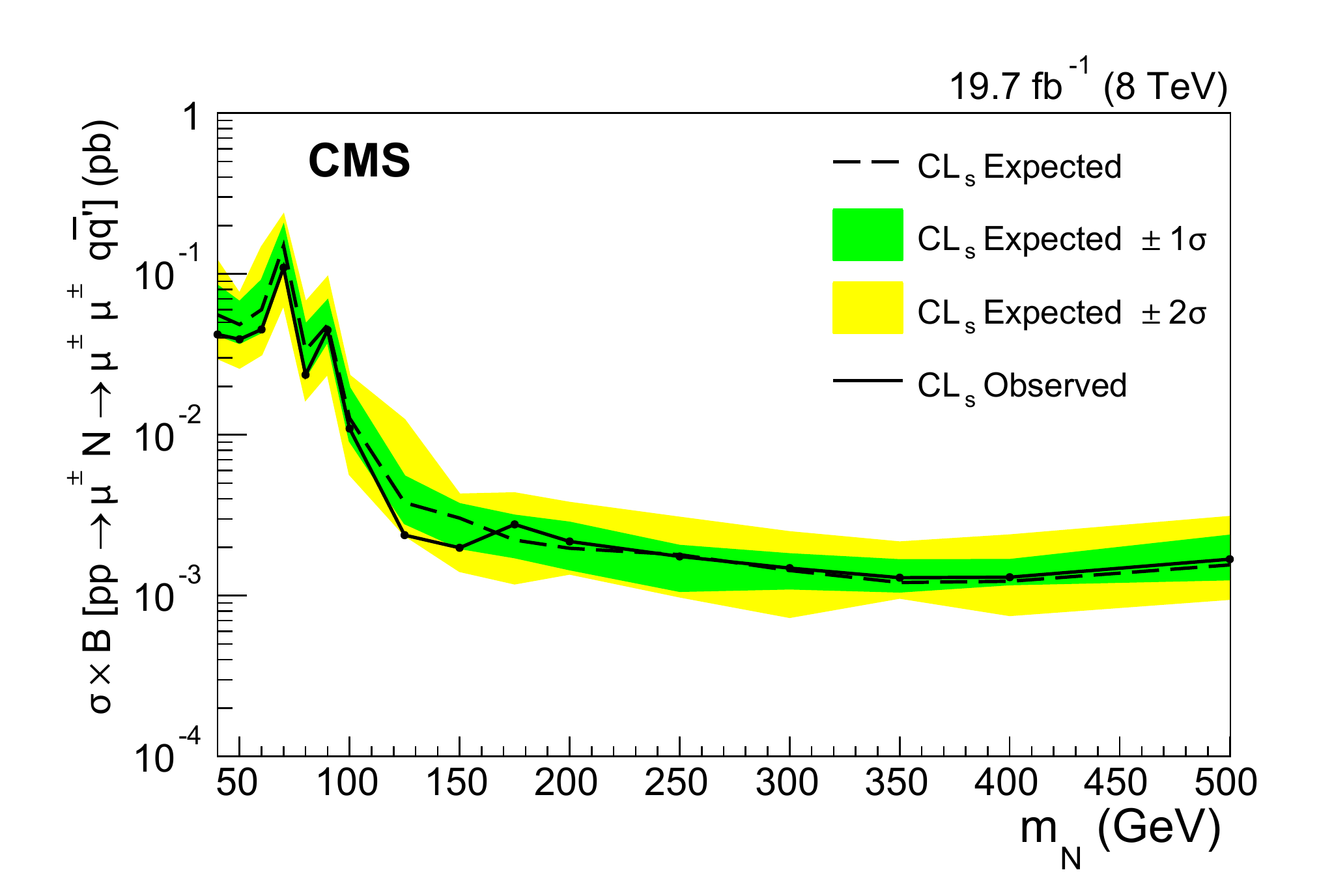}
  \caption{Exclusion region at 95\% CL in the cross section times branching fraction,
  as a function of the heavy Majorana neutrino mass.
  The long-dashed black curve is the expected upper limit, with one and two
  standard deviation bands shown in dark green and light yellow, respectively. The solid black curve is the
  observed upper limit.  The region above the exclusion
  curves is ruled out.
  }
\label{fig:xs_limit}
\end{figure}

We see no evidence for a significant excess in data beyond the backgrounds predicted from the SM
and we set 95\% confidence level (CL) exclusion limits  on the cross section times branching fraction for
$\Pp\Pp \to  \N \mu^\pm \to \mu^\pm \mu^\pm \PQq \PAQq^\prime$
as a function of $\mN$, using the CL$_s$ method~\cite{CLs1, CLs2, CLs3} based on the
event yields shown in Table~~\ref{table:yields}. We use Poisson distributions for the signal and
log-normal distributions for the nuisance  parameters. The limits obtained are shown in Fig.~\ref{fig:xs_limit}.
We do not consider the mass range below
$\mN = 40\GeV$ because of the very low selection efficiency for the signal in this mass region.

\begin{figure}[h*btp]\centering
  \includegraphics[width=\cmsFigWidth]{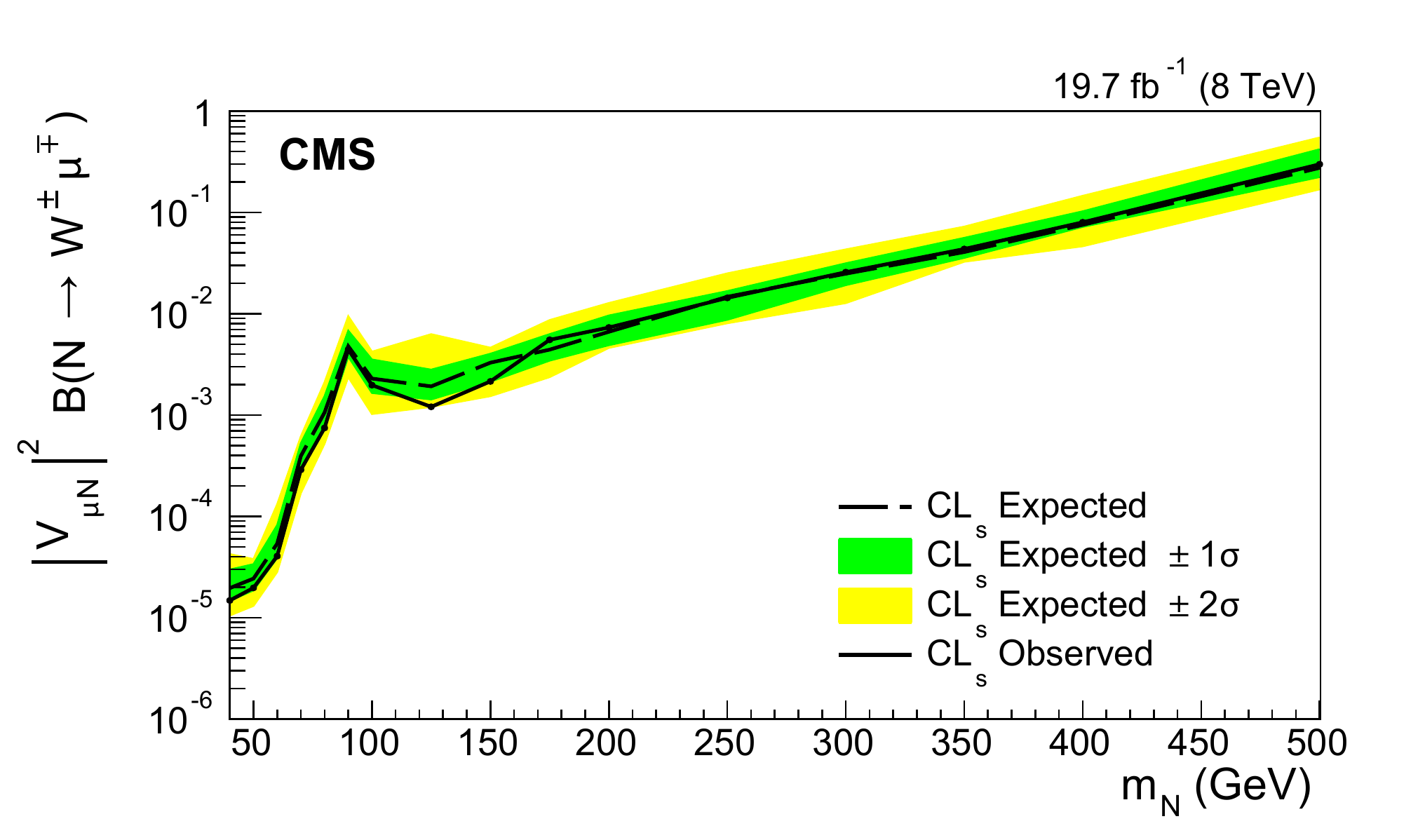}
  \includegraphics[width=\cmsFigWidth]{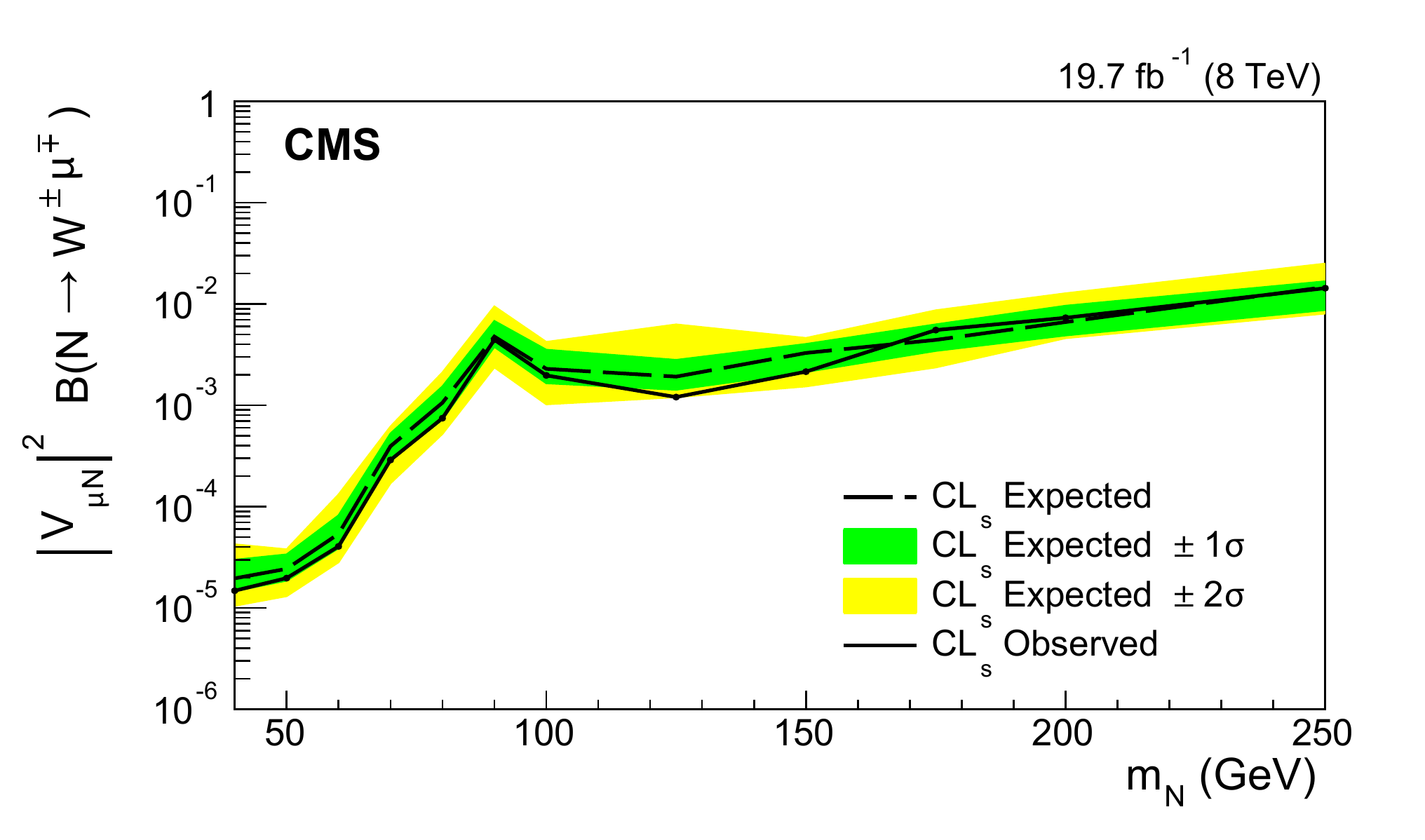}
  \caption{Exclusion region at 95\% CL in the square of the heavy Majorana neutrino mixing parameter
  times branching fraction of the N to a W boson and a muon,
  as a function of the heavy Majorana neutrino mass: ($\abs{\VmN}^2 \mathcal{B}(\N \to \PW^\pm \mu^\mp)$ vs. $\mN$).
  The long-dashed black curve is the expected upper limit, with one and two
  standard deviation bands shown in dark green and light yellow, respectively. The solid black curve is the
  observed upper limit. The regions above the exclusion
  curves are ruled out.
  The lower panel shows an expanded view of the region $40\GeV < \mN < 250\GeV$.
  }
\label{fig:excl}
\end{figure}

\begin{figure}[h*btp]\centering
  \includegraphics[width=\cmsFigWidth]{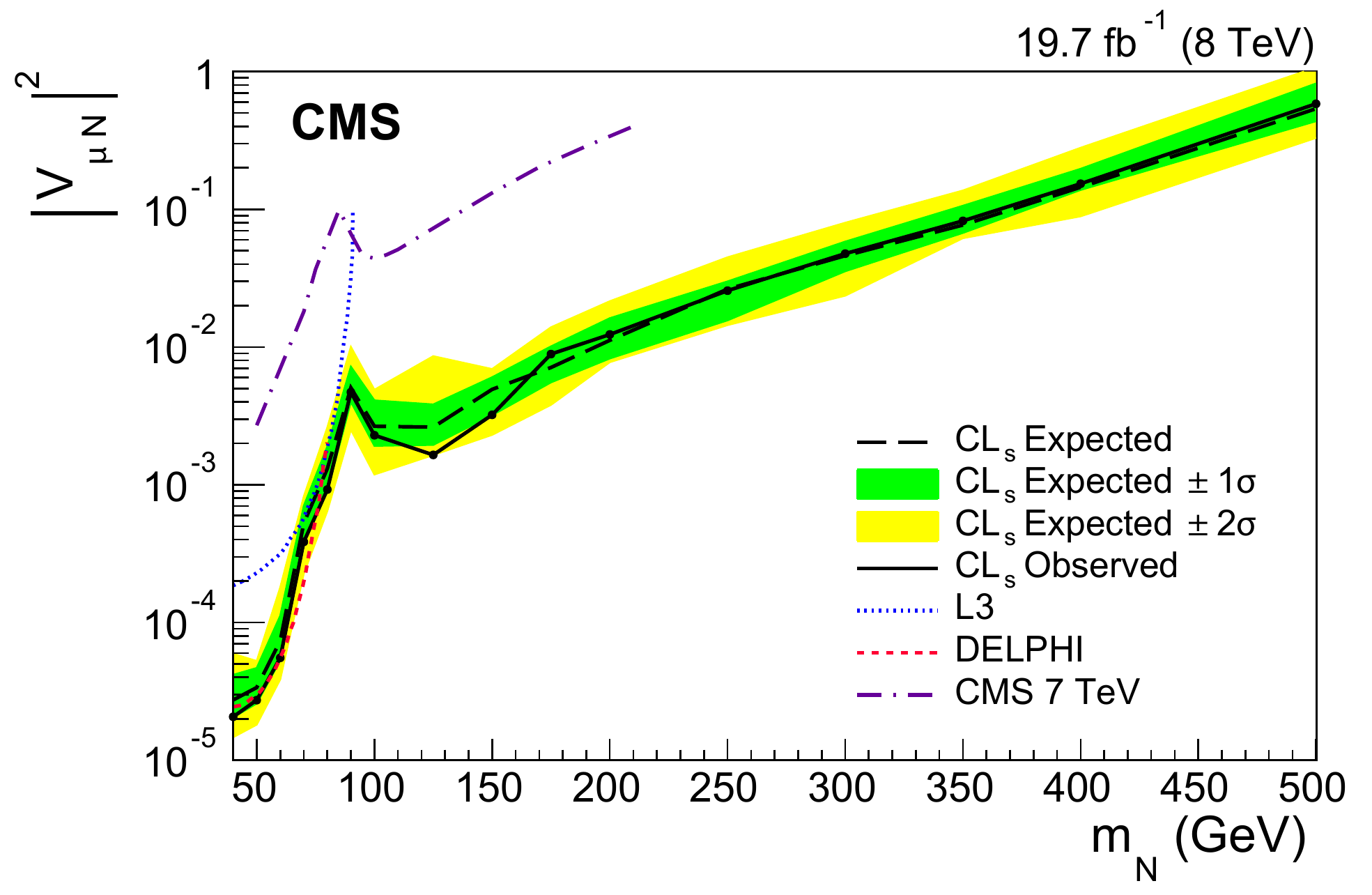}
  \includegraphics[width=\cmsFigWidth]{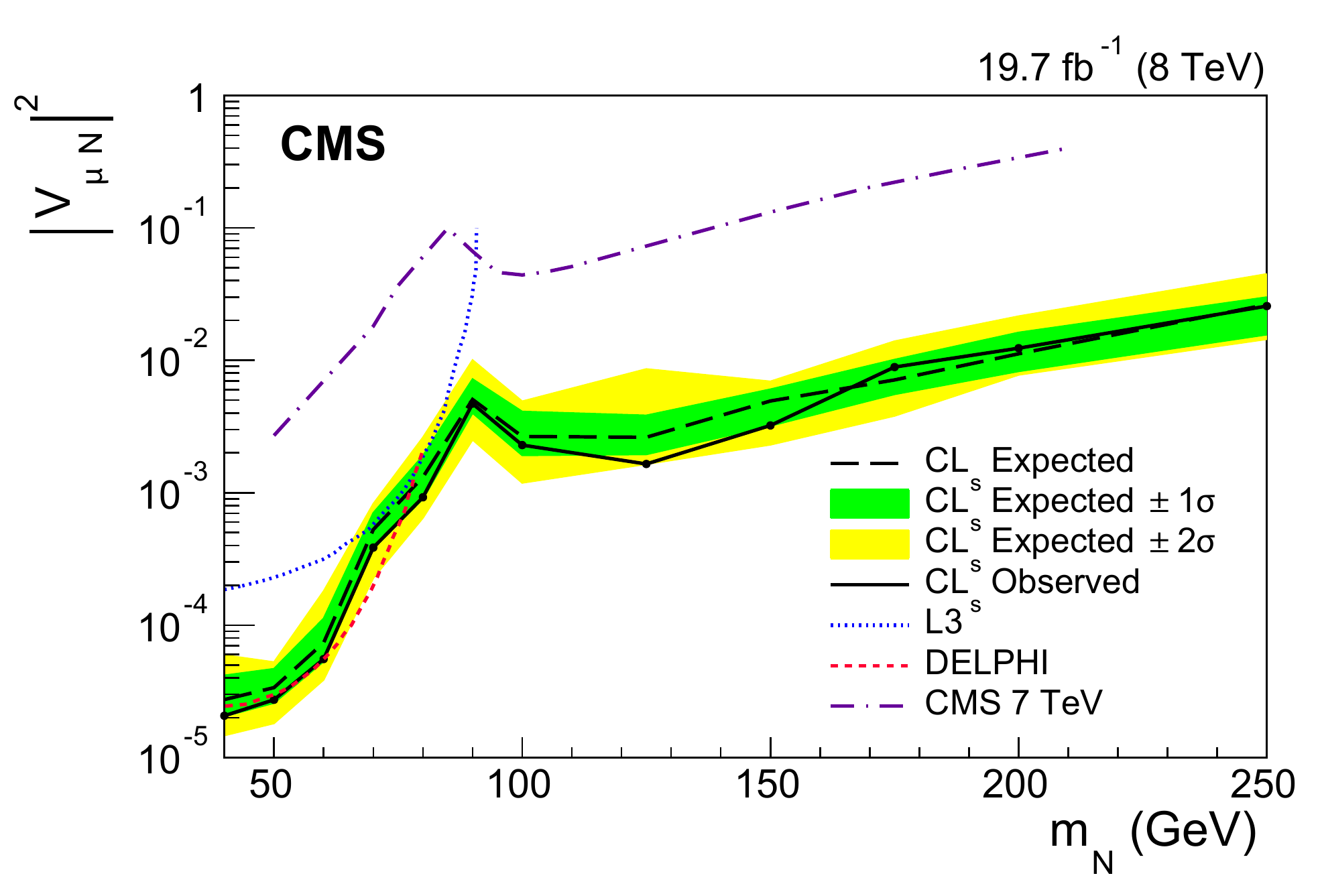}
  \caption{Exclusion region at 95\% CL in the square of the heavy Majorana neutrino mixing parameter
  as a function of the heavy Majorana neutrino mass: ($\abs{\VmN}^2$ vs. $\mN$).
  The long-dashed black curve is the expected upper limit, with one and two
  standard deviation bands shown in dark green and light yellow, respectively. The solid black curve is the
  observed upper limit.
  Also shown are the upper limits from other direct searches:  L3~\cite{l3}, DELPHI~\cite{delphi}, and the upper limits from
  CMS obtained with the 2011 LHC data at $\sqrt{s} = 7\TeV$~\cite{CMS_NR2011}. The regions above the exclusion
  curves are ruled out.
  The lower panel shows an expanded view of the region $40\GeV < \mN < 250\GeV$.
  }
\label{fig:excl2}
\end{figure}

The behavior of the limit around $\mN = 80\GeV$ is caused by the fact that as the heavy Majorana neutrino gets close to
the W-boson mass from below or above, the muon produced together with the N or the muon from the N decay have low $\pt$, respectively.
We also set limits on the square of the heavy Majorana neutrino mixing parameter times branching fraction of the N to a W boson and a muon, 
$\abs{\VmN}^2 B(\N \to \PW^\pm \mu^\mp)$, as a function of $\mN$.
Figure~\ref{fig:excl} shows the resulting upper limits
on $\abs{\VmN}^2  \mathcal{B}(\N \to \PW^\pm \mu^\mp)$ as a function of $\mN$.
Assuming the theoretical prediction for the branching fraction for $\N \to \PW^\pm \mu^\mp$, we can extract limits on 
$\abs{\VmN}^2$. These limits are shown in Fig.~\ref{fig:excl2}.


\section{Summary}
A search for heavy Majorana neutrinos in
$\mu^\pm \mu^\pm \mathrm{j j}$ events has been performed in
proton-proton collisions at a center-of-mass energy of 8\TeV,
using a data set corresponding to an integrated luminosity of
19.7\fbinv. No excess of events beyond the
standard model background prediction is found. Upper limits at 95\% CL are set on $\abs{\VmN}^2$,  as a
function of heavy Majorana neutrino mass $\mN$, where $\VmN$
is the mixing element of the heavy neutrino N with the standard model muon neutrino.

These are the first direct upper limits on the heavy Majorana neutrino mixing for $\mN > 200$\GeV.
The limits for \mN between 45\GeV and 90\GeV are comparable to the best limits from LEP~\cite{l3, delphi}, 
which are derived from $\cPZ \to \nu_\ell \N$ and are restricted to masses below approximately 90\GeV.
The limits reported here extend well beyond this mass and are significantly better than previous LHC limits
obtained at a center-of-mass energy of 7\TeV. For $\mN = 90$\GeV we find $\abs{\VmN}^2 < 0.00470$.
Furthermore, the direct limits on $\abs{\VmN}$ have been extended from
$\mN \approx 200\GeV$ up to $\mN \approx 500\GeV$.
At $\mN = 200$\GeV  the limit is $\abs{\VmN}^2 < 0.0123$, and at
$\mN = 500$\GeV  the limit is $\abs{\VmN}^2 < 0.583$.

\begin{acknowledgments}
We would like to thank Juan Antonio Aguliar-Saavedra, Francisco del Aguila, and Tao Han for assistance with Monte Carlo generators 
and calculations of the heavy neutrino branching fractions, as well as for advice on the theoretical interpretations of our results.

We congratulate our colleagues in the CERN accelerator departments for the excellent performance of the LHC and thank the technical and administrative staffs at CERN and at other CMS institutes for their contributions to the success of the CMS effort. In addition, we gratefully acknowledge the computing centers and personnel of the Worldwide LHC Computing Grid for delivering so effectively the computing infrastructure essential to our analyses. Finally, we acknowledge the enduring support for the construction and operation of the LHC and the CMS detector provided by the following funding agencies: BMWFW and FWF (Austria); FNRS and FWO (Belgium); CNPq, CAPES, FAPERJ, and FAPESP (Brazil); MES (Bulgaria); CERN; CAS, MoST, and NSFC (China); COLCIENCIAS (Colombia); MSES and CSF (Croatia); RPF (Cyprus); MoER, ERC IUT and ERDF (Estonia); Academy of Finland, MEC, and HIP (Finland); CEA and CNRS/IN2P3 (France); BMBF, DFG, and HGF (Germany); GSRT (Greece); OTKA and NIH (Hungary); DAE and DST (India); IPM (Iran); SFI (Ireland); INFN (Italy); MSIP and NRF (Republic of Korea); LAS (Lithuania); MOE and UM (Malaysia); CINVESTAV, CONACYT, SEP, and UASLP-FAI (Mexico); MBIE (New Zealand); PAEC (Pakistan); MSHE and NSC (Poland); FCT (Portugal); JINR (Dubna); MON, RosAtom, RAS and RFBR (Russia); MESTD (Serbia); SEIDI and CPAN (Spain); Swiss Funding Agencies (Switzerland); MST (Taipei); ThEPCenter, IPST, STAR and NSTDA (Thailand); TUBITAK and TAEK (Turkey); NASU and SFFR (Ukraine); STFC (United Kingdom); DOE and NSF (USA).

Individuals have received support from the Marie-Curie program and the European Research Council and EPLANET (European Union); the Leventis Foundation; the A. P. Sloan Foundation; the Alexander von Humboldt Foundation; the Belgian Federal Science Policy Office; the Fonds pour la Formation \`a la Recherche dans l'Industrie et dans l'Agriculture (FRIA-Belgium); the Agentschap voor Innovatie door Wetenschap en Technologie (IWT-Belgium); the Ministry of Education, Youth and Sports (MEYS) of the Czech Republic; the Council of Science and Industrial Research, India; the HOMING PLUS program of Foundation for Polish Science, cofinanced from European Union, Regional Development Fund; the Compagnia di San Paolo (Torino); the Consorzio per la Fisica (Trieste); MIUR project 20108T4XTM (Italy); the Thalis and Aristeia programs cofinanced by EU-ESF and the Greek NSRF; and the National Priorities Research Program by Qatar National Research Fund.
\end{acknowledgments}
\ifthenelse{\boolean{cms@external}}{}{\clearpage}
\bibliography{auto_generated}

\cleardoublepage \appendix\section{The CMS Collaboration \label{app:collab}}\begin{sloppypar}\hyphenpenalty=5000\widowpenalty=500\clubpenalty=5000\textbf{Yerevan Physics Institute,  Yerevan,  Armenia}\\*[0pt]
V.~Khachatryan, A.M.~Sirunyan, A.~Tumasyan
\vskip\cmsinstskip
\textbf{Institut f\"{u}r Hochenergiephysik der OeAW,  Wien,  Austria}\\*[0pt]
W.~Adam, T.~Bergauer, M.~Dragicevic, J.~Er\"{o}, M.~Friedl, R.~Fr\"{u}hwirth\cmsAuthorMark{1}, V.M.~Ghete, C.~Hartl, N.~H\"{o}rmann, J.~Hrubec, M.~Jeitler\cmsAuthorMark{1}, W.~Kiesenhofer, V.~Kn\"{u}nz, M.~Krammer\cmsAuthorMark{1}, I.~Kr\"{a}tschmer, D.~Liko, I.~Mikulec, D.~Rabady\cmsAuthorMark{2}, B.~Rahbaran, H.~Rohringer, R.~Sch\"{o}fbeck, J.~Strauss, W.~Treberer-Treberspurg, W.~Waltenberger, C.-E.~Wulz\cmsAuthorMark{1}
% ...
\vskip\cmsinstskip
% ...
1:~~Also at Vienna University of Technology, Vienna, Austria\\
2:~~Also at CERN, European Organization for Nuclear Research, Geneva, Switzerland\\
% ...

\end{sloppypar}
\end{document}